\documentstyle[preprint,aps,epsf]{revtex}

\newcommand{\slashed}[1]{\rlap{$#1$}/}
\newcommand{\slashp}{\mbox{$\not \hspace*{-1.10mm} p$}}

\newcommand{\slashq}{\mbox{$\not \hspace*{-1.10mm} q$}}

\newcommand{\UMLAUT}[1]{\"{#1}}
\newcommand{\GeV}{\mbox{\rm GeV}}

\begin{document}

\draft

\title{Two--photon processes of pseudoscalar mesons \\
in a Bethe--Salpeter approach}

\author{Dalibor Kekez}
\address{\footnotesize Rudjer Bo\v{s}kovi\'{c} Institute,
         P.O.B. 1016, 10001 Zagreb, Croatia}
\author{Dubravko Klabu\v{c}ar}
\address{\footnotesize Department of Physics, Faculty of Science, \\
        Zagreb University, P.O.B. 162, 10001 Zagreb, Croatia}

\maketitle

\begin{abstract}
We evaluate the $\pi^0\gamma^\star\to\gamma$ transition form factor 
and $\gamma\gamma$ decay widths for
$\pi^0, \eta_c$ and $\eta_b$ treated as $q\bar{q}$ bound 
states in the Bethe--Salpeter formalism,
incorporating the dynamical chiral symmetry breaking and the
Goldstone nature of the pion. 
In the chiral limit, the Abelian axial anomaly is 
incorporated analytically in our coupled Schwinger--Dyson
and Bethe--Salpeter approach,
which is also capable of quantitatively describing
systems of heavy quarks, concretely $\eta_c$
and possibly $\eta_b$. 

\vspace{1mm}
\noindent{\it PACS:} 11.10.St; 13.40. --f; 14.40.Aq; 14.40Gx \\
{\it Keywords:} Non-perturbative QCD; Hadronic structure; 
Schwinger--Dyson and Bethe--Salpeter equations; Axial anomaly;
Electromagnetic processes
\end{abstract}
\pacs{}


\noindent
{\bf 1.}
Recently, Bando {\it et al.} \cite{bando94} and Roberts \cite{Roberts} have 
demonstrated how the Adler-Bell-Jackiw axial anomaly can be incorporated
in the framework of Schwinger--Dyson (SD) and Bethe--Salpeter 
(BS) equations, reproducing 
(in the chiral limit)
the famous anomaly result for $\pi^0\to\gamma\gamma$ analytically.
This was also extended \cite{Frank+al} to the off--shell
case $\gamma^\star\pi^0\to\gamma$, a process of renewed 
experimental interest because of the plans to measure it
with high precision at CEBAF (see refs. in \cite{Frank+al,Burden+al}). 

%
%
%
Refs. \cite{Roberts,Frank+al} avoided solving the SD equation for 
the dressed quark propagator $S$ by using an {\it Ansatz} quark 
propagator. Then, thanks to working in the chiral limit,
they also automatically obtained the solution of the BS equation:
in this limit,
when the chiral symmetry is not broken explicitly but 
dynamically, and when pions must consequently appear as Goldstone bosons, 
the solution for the pion
bound-state vertex $\Gamma_\pi$, corresponding to the Goldstone pion,
is given (see Eq. (\ref{ChLimSol}), and, {\it e.g.,} \cite{Miransky}) by 
the dressed quark propagator $S(q)$: 

	\begin{equation}
	S^{-1}(q)
	=
	A(q^2)\slashed{q} - B(q^2)~.
	\label{quark_propagator}
	\end{equation}

\noindent Concretely, \cite{Roberts,Frank+al} used the solution 
for $\Gamma_\pi$ that is of zeroth order in the pion momentum $p$,
and which
fully saturates the Adler-Bell-Jackiw axial anomaly \cite{bando94,Roberts}.
It is proportional to $B(q^2)$ when (\ref{quark_propagator})
is the propagator of a massless quark, whereas
its normalization is given \cite{JJ} by the pion decay constant $f_\pi$:
        \begin{equation}
\Gamma_\pi(q;p^2\! =\! M_\pi^2\! =\! 0)\!
 =\, 2 \, \gamma_5 B(q^2)_{m=0}/f_\pi,    \label{ChLimSol}
        \end{equation}
\noindent (where the flavour structure of pions has been suppressed).
Thanks to using the 
chiral-limit solution (\ref{ChLimSol}) (and to satisfying 
the electromagnetic Ward-Takahashi identity), \cite{bando94,Roberts}
{\it analytically} reproduced
the famous axial-anomaly
result {\it independently} of what precise {\it Ansatz}
has been used \cite{Roberts}
for $S(q)$ and, consequently, independently  of what 
interaction $G^{\mu\nu}(k)$ has formed the $\pi^0$ bound state.

In the present paper we argue that in the light
of the recent experimental results on
$\eta_c\to\gamma\gamma$ from CLEO \cite{CLEO1995}, elements of
these treatments \cite{bando94,Roberts,Frank+al}, concerning the
electromagnetic interactions of dressed quarks, appear
essential also for the understanding of the electromagnetic processes
of mesons in a totally different regime, far away from the chiral limit.

However, the above--described scheme which avoids solving BS equations 
\cite{Roberts,Frank+al}, is restricted to
the chiral limit and very close to it. (Already when strange quarks
are present, (\ref{ChLimSol}) can be regarded only as an ``exploratory"
\cite{Burden+al} expression.) This scheme obviously must be
abandoned 
in the case of $c$-- and $b$--quarks, where the whole concept of the
chiral limit is of course useless even qualitatively.
Away from this limit, one must confront solving the pertinent 
bound--state equation, 
which is  determined by the interaction between quarks.
On the other hand, one should also have the axial anomaly incorporated
correctly. This consistency requirement is satisfied by  
the coupled SD-BS approach to $q\bar q$ bound states
presented by Jain and Munczek \cite{jain91,munczek92,jain93b}
(and reviewed in a broader SD-BS context by \cite{Miransky,RW}),
because this treatment in the chiral limit yields pions as
Goldstone bosons of dynamical chiral symmetry breaking (D$\chi$SB).  
On the other hand, it has also successfully reproduced 
\cite{jain91,munczek92,jain93b}
almost the whole spectrum of meson masses, 
including those in the heavy-quark regime,
and also the leptonic decay constants ($f_P$)
of pseudoscalar mesons ($P$).
We are therefore motivated to use this so far
successful approach for calculating other quantities.
In this paper we present the calculation of the $\pi^0, \eta_c$, 
and $\eta_b \rightarrow \gamma\gamma$ decay widths, as well as
of the $\gamma^\star\pi^0$-to-$\gamma$ transition form factor.
(Of course, in the case of $\eta_b$ we do not question
the accuracy of the non--relativistic description, but want
to see how well our approach does once the experimental data
on $\eta_b$ are obtained.)

\vspace{5mm}


\noindent
{\bf 2.}
To define the interaction kernel for the SD and BS equations,
\cite{jain91,munczek92,jain93b} use the modeled Landau-gauge gluon
propagator given 
as the sum of the following two parts:
\begin{equation}
             G^{\mu\nu}(k) = [ G_{UV}(-k^2) + G_{IR}(-k^2) ]
        ( g^{\mu\nu} - \frac{k^\mu k^\nu}{k^2} )~,
        \label{gluon_propagator}
        \end{equation}
%
%
%
%
%
\noindent which define the separation of the propagator into:
a) the well--known perturbative part, correctly
reproducing the ultraviolet (UV) asymptotic behaviour unambiguously
required by QCD in its high--energy, perturbative regime, and
b) the nonperturbative part, which should describe the infrared
(IR) behaviour. The infrared behaviour of QCD is however still 
not well understood,
so it is only this latter, nonperturbative part that
is in fact modeled. 
{}From the renormalization group, in the spacelike region ($Q^2 = -k^2$),

	\begin{equation}
	G_{UV}(Q^2)
	=
                  \frac{16\pi}{3}\frac{\alpha_s(Q^2)}{Q^2}
	\approx
\frac{{\frac{16\pi^2}{3}} d}{Q^2 \ln(x_0+\frac{Q^2}{\Lambda_{QCD}^2})}
		{\huge \{}
			1
			+
                 b \, \frac{\ln[\ln(x_0+ \frac{Q^2}{\Lambda_{QCD}^2})]}
			       {\ln(x_0+ \frac{Q^2}{\Lambda_{QCD}^2})}
		{\huge \}}~,
\label{gluon_UV}
	\end{equation}

\noindent where the two--loop asymptotic expression for $\alpha_s(Q^2)$
is employed, and where $d=12/(33-2N_f)$, $b=2\beta_2/\beta_1^2=
2(19N_f/12 -51/4)/(N_f/3 -11/2)^2$,  and $N_f$ is the number of quark
flavours. Following \cite{jain93b}, we set $N_f=5$,
$\Lambda_{QCD}=228\,\mbox{\rm MeV}$, and $x_0=10$.
For the modeled, IR part of the gluon propagator, we choose
$G_{IR}$ also from Ref.~\cite{jain93b}:

	\begin{equation}
	G_{IR}(Q^2)
	=
                   {\frac{16\pi^2}{3}} \,a\,Q^2 e^{-\mu Q^2},
	\label{gluon_IR}
	\end{equation}

\noindent  
with their \cite{jain93b} parameters
$a=(0.387\,\GeV)^{-4}$ and $\mu=(0.510\,\GeV)^{-2}$
adopted throughout.

We obtain the dressed quark propagators (\ref{quark_propagator})
for various flavours
by solving the SD equation in the ladder (or ``rainbow"
\cite{RW}) approximation ({\it i.e.}, with bare quark-gluon vertices):

        \begin{equation}
        S^{-1}(q) = \slashed{q} - \widetilde{m} - i
\int \frac{d^4k}{(2\pi)^4}
                    \gamma^\mu S(k) \gamma^\nu G_{\mu\nu}(q-k)~,
       \label{SD-equation}
        \end{equation}

\noindent where $\widetilde{m}$ is the bare
mass term of the pertinent quark flavour,
breaking the chiral symmetry explicitly.
({\it I.e.}, we do not use any 
{\it Ans\" atze} for $A(q^2)$ or $B(q^2)$.)
The case $\widetilde{m}=0$ corresponds to the chiral limit where
the current quark mass $m=0$, and where the constituent quark mass
$B(0)/A(0)$ stems exclusively from D$\chi$SB.
For $u$ and $d$-quarks, assumed massless, solving of (\ref{SD-equation})
leads to $B(0)/A(0)=375$ MeV for the parameters from \cite{jain93b}.
When $\widetilde{m}\neq 0$, the SD equation (\ref{SD-equation})
must be regularised by a UV cutoff $\Lambda$ \cite{munczek92,jain93b},
and $\widetilde{m}$ is in fact a cutoff--dependent quantity.
In \cite{jain93b}, $\Lambda=134$ GeV, while $\widetilde{m}(\Lambda^2)$ 
is $680$ MeV for $c$--quarks, and $3.3$ GeV for $b$--quarks.
This gives us the constituent mass $B(0)/A(0)$ 
of $1.54$ GeV for the
$c$--quarks, and $4.77$ GeV for the $b$--quarks.
In the chiral limit, 
Eqs. (\ref{quark_propagator})--(\ref{ChLimSol}) reflect the fact that 
solving of (\ref{SD-equation}) for $S(q)$ with $\widetilde{m}=0$ is already 
sufficient to give us the Goldstone pion bound-state vertex
(\ref{ChLimSol})
saturating the anomalous $\pi^0\to\gamma\gamma$ decay, and that
is how our $\Gamma_\pi$ is obtained.
Of course, we cannot avoid solving the BS equation
like this for heavier $q\bar q$ pseudoscalars $P$, such as
$P=\eta_c$, $\eta_b$.
Their bound-state vertices $\Gamma_{P}$ 
must be obtained by explicit solving of

        \begin{equation}
                \Gamma_{P}(q,p) 
                = i 
                \int \frac{d^4q^\prime}{(2\pi)^4}
                \gamma^\mu S(q^\prime + \frac{p}{2})
      \Gamma_{P}(q^\prime,p)   S(q^\prime - \frac{p}{2})
                \gamma^\nu
                G_{\mu\nu}(q-q^\prime)~,
        \label{BSE}
        \end{equation}

\noindent the homogeneous BS equation again in the 
(``generalized" \cite{Roberts}) ladder approximation,
consistently with (\ref{SD-equation}). 
(For $\eta$ and $\eta^\prime$, their  mixing complicates the situation,
so we do not solve for $P=\eta, \eta^\prime$ at this point.)
Note that $S$ is the quark propagator 
obtained by solving the SD equation (\ref{SD-equation}) with the same
gluon propagator $G_{\mu\nu}$. 
For pseudoscalar mesons,
the complete decomposition of the BS bound state vertex $\Gamma_P$
in terms of the scalar functions $\Gamma^P_i$ is:

        \begin{equation}
\Gamma_P(q,p)=\gamma_5 \left\{\, \Gamma^P_0(q,p) + \slashp \, \Gamma^P_1(q,p)
  + \slashq \, \Gamma^P_2(q,p) + [\slashp,\slashq]\, \Gamma^P_3(q,p)\, \right\}.
        \end{equation}  

\noindent (The flavour indices are again suppressed.)
The BS equation (\ref{BSE}) leads
to a coupled set of integral equations for the functions
$\Gamma^P_i$ $(i=0,...,3)$,
which is most easily solved numerically in the Euclidean space. 
Solving for $\Gamma^{\eta_c}_i$, we also get the $\eta_c$ mass
$M_{\eta_c}=2.875$ GeV, whereas experimentally $M_{\eta_c}=2.979$
GeV. Nevertheless, this is just one example of
the successful reproduction of heavy-meson masses  
in this approach \cite{jain91,munczek92,jain93b}.
For $\eta_b$, where there are no experimental results yet, we predict
$M_{\eta_b}=9.463$ GeV.

\vspace{5mm}


\noindent
{\bf 3.}
The transition matrix element for the processes $\pi^0, \eta_c, \eta_b
\rightarrow \gamma\gamma$, and $\gamma^\star\pi^0 \rightarrow \gamma$,
is given by the famous triangle graph coupling two photons 
(with momenta $k$ and $k^\prime$) to a neutral pseudoscalar $P$. 
Apart from the momentum-conserving delta-function ensuring
$p=k+k^\prime$,
and the photon polarization vectors $\varepsilon^{\mu}(k,\lambda),
\varepsilon^{\nu}(k^\prime,\lambda^\prime)$, it is
in essence equivalent to a
tensor amplitude $T_P^{\mu\nu}(k,k^\prime)$,
which is the time ordered product of
two electromagnetic currents of quarks, evaluated 
between a pseudoscalar state $|P\rangle$ and the vacuum
state $\langle 0|$.
By symmetry arguments, it can be written as \cite{itzykson80}
        $T_P^{\mu\nu}(k,k^\prime)
        =
        \varepsilon^{\alpha\beta\mu\nu} k_\alpha k^\prime_\beta
        T_P(k^2,k^{\prime 2})~,
        $
where $T_P(k^2,k^{\prime 2})$ is a scalar.
When both of the photons are 
on their mass shells, $k^2=0$ and $k^{\prime 2}=0$,
the decay width is 
        $
        \Gamma(P\to\gamma\gamma)
        =
        \pi\alpha^2 M_P^3
     \, |T_P(0,0)|^2 / 4~. \,$ 
$(P = \pi^0, \eta_c, \eta_b.)$

On the other hand, for the $\gamma^\star\pi^0\to\gamma$ transition,
only the final photon $\gamma$ is necessarily on the mass shell, $k^2=0$,
whereas $k^{\prime 2}=-Q^2\neq 0$ for the virtual photon $\gamma^\star$.
The $\gamma^\star\pi^0\to\gamma$ transition form factor $F(Q^2)$ on
the $\pi^0$ mass shell is then defined as

	\begin{equation}
	F(Q^2)
	=
        T_{\pi^0}(0,-Q^2)/T_{\pi^0} (0,0)~.
        \end{equation}	

How 
to calculate $T_P$, or, equivalently, $T_P^{\mu\nu}$?
The framework for incorporating electromagnetic interactions
in the context of BS bound states of dressed quarks, advocated 
by (for example) \cite{bando94,Roberts,Frank+al,Burden+al}
and called the generalized impulse approximation (GIA) by
\cite{Roberts,Frank+al,Burden+al}, is probably the
simplest framework consistent with our coupled SD--BS approach. 
This means that in the triangle graph we use our dynamically
{\it dressed} quark propagator $S(q)$, Eq.~(\ref{quark_propagator}),
and the pseudoscalar BS bound--state vertex $\Gamma_P(q,p)$
instead of the bare $\gamma_5$ vertex. Another essential 
element of the GIA is to use an appropriately dressed {\it electromagnetic}
vertex $\Gamma^\mu(q^\prime,q)$, which satisfies the vector Ward--Takahashi
identity (WTI),

        \begin{equation}
        (q^\prime-q)_\mu \Gamma^\mu(q^\prime,q) =
                S^{-1}(q^\prime) - S^{-1}(q)~.
        \label{WTI-v}
        \end{equation}

\noindent Our 
dressed 
quark propagator
contains the momentum-dependent functions $A(q^2)$ and $B(q^2)$,
so that the bare vertex $\gamma^\mu$ violates (\ref{WTI-v}),
implying the nonconservation of the electromagnetic vector current
and of the electric charge. Therefore, as in GIA, we must also use 
such a dressed, WTI-preserving quark--photon vertex.
Solving the pertinent SD equation for the true, 
dressed quark-photon vertex $\Gamma^\mu$ 
is a difficult problem which has only recently 
begun to be addressed \cite{F}. Therefore, it is
customary to use realistic {\it Ans\"{a}tze} \cite{RW}.
Motivated by its successes in \cite{Roberts,Frank+al},
we choose the Ball--Chiu (BC) ~\cite{BC} vertex for
$\Gamma^\mu(q^\prime,q)$:

        \begin{eqnarray}
        \Gamma^\mu(q^\prime,q) =
        A_{\bf +}(q^{\prime 2},q^2)
       \frac{\gamma^\mu}{\textstyle 2}
        + \frac{\textstyle (q^\prime+q)^\mu }
               {\textstyle (q^{\prime 2} - q^2) }
        \{A_{\bf -}(q^{\prime 2},q^2)
        \frac{\textstyle (\slashed{q}^\prime + \slashed{q}) }{\textstyle 2}
         - B_{\bf -}(q^{\prime 2},q^2) \}~,
        \label{BC-vertex}
        \end{eqnarray}

\noindent where
$H_{\bf \pm}(q^{\prime 2},q^2)\equiv [ H(q^{\prime 2}) \pm H(q^2) ]$,
for $H = A$ or $B$. 
Obviously, this {\it Ansatz} does not introduce any
new parameters as it is completely determined 
for each flavour by the pertinent solution for the quark
propagator (\ref{quark_propagator}). Its four chief virtues, however,
are {\it (i)} that it satisfies the Ward--Takahashi identity (\ref{WTI-v}),
{\it (ii)} that it reduces to the bare vertex in the free-field limit
as must be in perturbation theory, {\it (iii)} that its transformation
properties under
Lorentz transformations and charge conjugation are the same as for the
bare vertex, and {\it (iv)} it has no kinematic singularities.

It is important to note that the correct
axial-anomaly result cannot be obtained \cite{Roberts}
analytically in the chiral limit unless 
a quark--photon--quark ($qq\gamma$) vertex
that satisfies the Ward-Takahashi identity 
is used (even if D$\chi$SB {\it is} employed and the
pion {\it does} appear as a Goldstone boson, as
comparison with \cite{horvat91,horvat95} shows).

In the case of $\pi^0$ for example, the GIA yields 
the amplitude $T_P^{\mu\nu}(k,k^\prime)$:

\begin{displaymath}
        T_{\pi^0}^{\mu\nu}(k,k^\prime)
        =
        -
        N_c \,
        \frac{Q_u^2-Q_d^2}{2}
        \int\frac{d^4q}{(2\pi)^4} \mbox{\rm tr} \{
        \Gamma^\mu(q-\frac{p}{2},k+q-\frac{p}{2})
        S(k+q-\frac{p}{2})
\end{displaymath}
\begin{equation}
	  \qquad
        \times
        \Gamma^\nu(k+q-\frac{p}{2},q+\frac{p}{2})
        S(q+\frac{p}{2})
        \Gamma_{\pi^0}(q,p)
        S(q-\frac{p}{2}) \}
        +
        (k\leftrightarrow k^\prime,\mu\leftrightarrow\nu).
\label{Tmunu(2)}
\end{equation}

\noindent (The analogous expressions for 
$\eta_c$ and $\eta_b$, or any other neutral pseudoscalar,
are straightforward.)
The number of colours $N_c$ arose from the trace
over the colour indices. The $u$ and $d$ quark charges,
$Q_u=2/3$ and $Q_d=-1/3$, appeared from tracing the product
of the the quark charge operators $Q$ and $\tau_3/2$,
the flavour matrix appropriate for $\pi^0$:
${\rm tr}(Q^2\tau_3/2)=(Q_u^2-Q_d^2)/2$.

\vspace{5mm}


\noindent
{\bf 4.} 
Let us first 
discuss the $\pi^0\gamma^\star\to\gamma$ transition form factor
$F(Q^2)$ for space--like transferred momenta.
The existing CELLO data \cite{behrend91},
displayed (following \cite{Frank+al}) as $Q^2 F(Q^2)$ in Fig. 1, are
described well by a monopole curve with the slope at $Q^2 =0$
which is equivalent to the ``interaction size''
${\langle r^2_{\gamma\pi^0\gamma} \rangle}^{1/2}_{exp}=0.65 \pm 0.03$ fm
(defined in, {\it e.g.}, \cite{Frank+al} via
$\langle r^2_{\gamma\pi^0\gamma} \rangle = - 6 F^\prime(Q^2)_{Q^2=0}$).
This was reproduced in various models.
{\it E.g.}, in the relativistic constituent quark model with an 
{\it Ansatz} for the pion wave function, Jaus \cite{Jaus} obtained a 
slope parameter equivalent to 
${\langle r^2_{\gamma\pi^0\gamma} \rangle}^{1/2}=0.64$ fm,
although the transition amplitude at
$Q^2 =0$ ($\pi^0\to\gamma\gamma$ transition), is 20\% too small.
In a generalization of the Nambu-Jona-Lasinio model,
Ito {\it et al.} \cite{IBG} got $F(Q^2)$ in excellent
agreement with the data and with the monopole fit with
${\langle r^2_{\gamma\pi^0\gamma} \rangle}^{1/2}=0.65$ fm.
The 
amplitude obtained in the
chiral perturbation theory gave (for small $Q^2$'s) 
$F(Q^2) = 1 - \langle r^2_{\gamma\pi^0\gamma} \rangle Q^2/6$, with 
${\langle r^2_{\gamma\pi^0\gamma} \rangle}^{1/2}=0.28$ fm \cite{BBC},
but introduction of vector mesons increased it to
${\langle r^2_{\gamma\pi^0\gamma} \rangle}^{1/2}=0.64$ fm \cite{ABBC}.
Bando and Harada's \cite{BH} Breit-Wigner forms 
for the propagators of $\rho$ and $\omega$ resonances,
have infinite derivatives for $Q^2 \to 0_{+}$,
so that one cannot give ${\langle r^2_{\gamma\pi^0\gamma} \rangle}$,
although their $F(Q^2)$, when pure vector-meson dominance is assumed,
almost coincides with the monopole curve with 
${\langle r^2_{\gamma\pi^0\gamma} \rangle}^{1/2}=0.65$ fm.
(See Fig. 1.)
We also display the Brodsky--Lepage interpolation curve
\cite{BrodskyLepage} to the perturbative QCD factorization limit
$Q^2 F(Q^2)\rightarrow 8\pi^2 f_\pi = 0.67$ GeV$^2$.
However, our results (the solid line in Fig. 1) should in the 
first place be compared with 
$F(Q^2)$ of the closely related approach of Frank {\it et al.}
\cite{Frank+al} (the dashed line). Our $F(Q^2)$ is a little better,
but it is in fact relatively similar to the curve of \cite{Frank+al},
considering that our $B(q^2)$, obtained by solving of the SD
equation (\ref{SD-equation}) (in this case in the chiral limit),
is very different from the {\it Ansatz} for $B(q^2)$ used by
\cite{Frank+al}.
Also, our ${\langle r^2_{\gamma\pi^0\gamma} \rangle}^{1/2}=0.46$ fm
is practically the same as 0.47 fm of \cite{Frank+al},
so that the ``interaction size"
discriminates even less between our respective $B(q^2)$'s, 
which describe the structure of the pion (\ref{ChLimSol})
in the chiral limit. It seems, therefore, that the differences 
in the internal structure between our pion and that of \cite{Frank+al}
play a relatively small role for $\pi^0\gamma^\star\to\gamma$,
at least in the chiral limit.

For  the on-shell case $\pi^0\to\gamma\gamma$,
the dependence on the pion structure 
falls out completely in this limit. 
Namely, our framework is in the chiral limit equivalent 
to \cite{bando94,Roberts,Frank+al}, as demonstrated by
the fact that, 
in this limit, with the solution (\ref{ChLimSol}), we too can reproduce
the famous anomaly result $T_{\pi^0}(0,0)=1/4\pi^2 f_\pi$
analytically, in the closed form.
The decay width is then given by
$
\Gamma(\pi^0 \rightarrow \gamma\gamma) =
      (\alpha^2/64\pi^3)\, (M_\pi^3/f_\pi^2)
    $ \cite{itzykson80},
in excellent agreement with experiment -- see Table~I.
Being a consequence of respecting the WTI (\ref{WTI-v})
and incorporating D$\chi$SB,
this result is independent
of our concrete choice of the interaction kernel and the
resulting hadronic structure of $\pi^0$, {\it i.e.}, of $B(q^2)$.
This is as it should be,
because the axial anomaly, which dominates $\pi^0\to\gamma\gamma$,
is of course independent of the structure.
It is then not surprising that calculations
for $\pi^0\to\gamma\gamma$ which rely on the details
of the hadronic structure 
(instead on the axial anomaly)
have problems to describe this decay accurately.

Of course, the situation is very different for quark masses
of the order of $\Lambda_{QCD}$ and higher: only the numerical
evaluation of $T_P(0,0)$ is reliable in this regime.  
Moreover, the details of the chosen interaction kernel
and the resulting propagator functions $A(q^2)$ and $B(q^2)$,
as well as the bound state solutions, {\it do} matter in that regime
(which gets further and further away from the domination of the 
axial anomaly with growing quark masses, as demonstrated
vividly by the amplitude ratios in the last column of Table~I).
This is naturally the case with the heavy-quark composites 
$\eta_c$ and $\eta_b$. Their two--photon widths are also given in Table~I.
In the case of $\eta_b$, 
only predictions exist, as there are no experimental data yet. 
In the case of $\eta_c$, our $\gamma\gamma$ width 
is (without any adjustment of model parameters)
comfortably within the empirical error bars after 
the 1994  Particle Data Group average,
and its 1995 electronic update \cite{PDG95}, have been further
updated with the new, 1995 CLEO \cite{CLEO1995}
result of $(4.3\pm 1.0\pm 0.7\pm 1.4)$ keV.
As opposed to this, it seems that most of the other theoretical
approaches predict $\eta_c\to \gamma\gamma$ decay widths
that will be  either too large or too small after
inclusion of the 1995 CLEO \cite{CLEO1995} result in the
Particle Data Group average.

To elaborate on this, let us remark that the results for 
$\Gamma(\eta_c\to \gamma\gamma)$ in the constituent, nonrelativistic 
potential models, once in good agreement with experiment, 
have recently been shown \cite{AhmadyMendel} to 
rise to far too large values of $11.8\pm 0.8\pm 0.6$ keV
after the calculations have been {\it improved} by removing 
certain approximations. 
The estimates \cite{AhmadyMendel} of the relativistic corrections 
indicate that they can reduce the width back down to some 8.8 keV.
This is not enough to bring the constituent quark model 
in agreement with experiment, although the relativistic corrections
are substantial, corroborating 
the view that the relativistic approaches to bound states,
such as ours, retain their importance on the mass scale of the
$c$--quarks.
This view has recently been strongly advocated by a
broad overview \cite{Muenz} of the two--photon physics of 
mesonic $q\bar q$--composites in the BS approach
(but without D$\chi$SB
and without a dressed $qq\gamma$-vertex); it has concluded that
the relativistic effects are important for {\it any} two--photon
width, even for heavy quarkonia. It is interesting to note,
however, that such BS calculations,
typically yielding \cite{Muenz} $\Gamma(\eta_c\to \gamma\gamma)$
below 4 keV, are consistent only with the
1995 CLEO \cite{CLEO1995} result on $\Gamma^{exp}
(\eta_c\to \gamma\gamma)$, whereas they are too low for all other
\cite{PDG95} measurements,
and even for the newly lowered average of $6.2\pm 1.3$ keV
in Table~I. ({\it E.g.}, see \cite{Muenz}
and other related BS calculations discussed therein.
They underestimate 
$\gamma\gamma$-widths in the light sector even more.)

The results of our approach are higher and in agreement 
with the empirical decay widths.
Our results also reveal the importance of using appropriately 
dressed quark--photon vertices together with dressed quark 
propagators in yet another way, which has to do with
internal consistency. To demonstrate the consequences 
of violating the WTI (\ref{WTI-v}),
and thereby of non--conservation of electric charge,
we compare (in the fourth column of Table~I) the 
decay widths $\Gamma^{bare}{(P\to\gamma\gamma)}
$
obtained inconsistently, {\it i.e.}, in the coupled SD--BS approach 
\cite{jain93b} but using the bare electromagnetic vertex $\gamma_\mu$,
with our results $\Gamma(P\to\gamma\gamma)$ using
the WTI--preserving dressed vertex (\ref{BC-vertex}). 
While not so catastrophic as for the light $\pi^0$, the violation of the WTI
(\ref{WTI-v}) is still very bad for $\eta_c\to\gamma\gamma$, 
and very noticeable
even for $\eta_b$ composed of very heavy $b$--quarks, for which 
the dynamical dressing is not as important as in the light sector.

The reader should note that we have not done any
fine--tuning of the parameters, or of the gluon propagator
form (\ref{gluon_propagator})-(\ref{gluon_IR}) which we used; these are
the  propagator and the parameters of Ref.~\cite{jain93b}, which
achieved a broad fit to the meson spectrum and pseudoscalar decay constants.
Therefore, in the consistently applied generalized impulse approximation,
the coupled SD--BS approach genuinely, without fitting,
leads to the adequate amplitude strength, which is otherwise too low.
In other words, the BS approach \cite{jain91,munczek92,jain93b} 
which is consistent with the ideas of
\cite{bando94,Roberts,Frank+al} on dressed quark--photon interactions, 
seems to be able to describe electromagnetic 
processes well even in the heavy-quark sector without fine-tuning of 
model parameters.
On the other hand, since $\Gamma(\eta_c\to \gamma\gamma)$ {\it does}
depend on the interaction kernel
and the parameters determining the internal structure of $\eta_c$,
making the measurements of the processes such as 
$\eta_c\to \gamma\gamma$ more 
precise can, through our theoretical approach, contribute to
determining the nonperturbative gluon propagator more accurately.
In the first place, this pertains to
the infrared part, which is at present poorly known,
but is also of significance for the determination
of the QCD running coupling $\alpha_s$
(contained in (\ref{gluon_UV}), the ultraviolet part of our gluon
propagator),
for example, at the scale $m_c$ naturally sampled by the $\eta_c$ system.


\vspace{5mm}

{\small 
The authors acknowledge
the support of 
the Croatian Ministry of Science and
Technology under the contracts 1--03--233 and 1--03--068 as well as 
the EU contract CI1*--CT91--0893 (HSMU).
The authors thank P. Jain for very useful discussions. 
}


\begin{table}
\begin{tabular}{ccccc}
$P$ &  $\Gamma(P\to\gamma\gamma)$ & 
$\Gamma^{exp}(P\to\gamma\gamma)$ &
$\frac{\textstyle{\Gamma^{bare}{(P\to\gamma\gamma)}}}
{\textstyle{\Gamma(P\to\gamma\gamma)}}$& 
$ \frac{\textstyle{T_{\pi^0}(0,0)}}{\textstyle{T_P(0,0)}}$\\
\hline
$\pi^0$  & 7.7 & 7.74$\pm 0.56$ & 0.22 &$ 1 $ \\
$\eta_{c}$   & $5.3\times 10^{+3}$ &
$(6.2\pm 1.3)\times 10^{+3}$ & 0.43 & $ 3.95 $\\
$\eta_{b}$   & $0.15 \times 10^{+3}$ & ? & 0.73 & $133$ \\
\end{tabular}
\caption{Comparison of the calculated $\gamma\gamma$ decay widths
(in eV) of 
$\pi^0, \eta_c$ and $\eta_b$
with their average experimental widths, where we have
included the 1995 CLEO result \protect\cite{CLEO1995}. 
The widths $\Gamma^{bare}(P\to \gamma\gamma)$, 
obtained with the bare electromagnetic
vertices $\gamma_\mu$, are significantly reduced with respect to
$\Gamma(P\to \gamma\gamma)$ due to the non--conservation 
of charge even in the heavy--quark regime. 
At present, there are no experimental data on $\eta_b$,
so the calculated mass of $\eta_b$ had to be used in the
phase--space factors, unlike $\pi^0$ and $\eta_c$.}
\label{tab:resullts_01}
\end{table}


\newpage

\section*{Figure captions}

\begin{itemize}

%
	\item[{\bf Fig.~1:}]
$\gamma^\star\pi^0\to\gamma$ form factor. The solid line represents
our results. The dashed line represents the results
of Ref.~\protect\cite{Frank+al}. The dash--dotted line is the Brodsky--Lepage
interpolation \protect\cite{BrodskyLepage}. The line of open circles is 
the curve of \cite{BH}, and open squares form the monopole curve 
corresponding to ${\langle r^2_{\gamma\pi^0\gamma} \rangle}^{1/2}=0.65$
fm. Experimental points are the results of the CELLO collaboration 
\protect\cite{behrend91}.

\end{itemize}


\newpage

\vspace*{2cm}
\epsfxsize = 18 cm \epsfbox{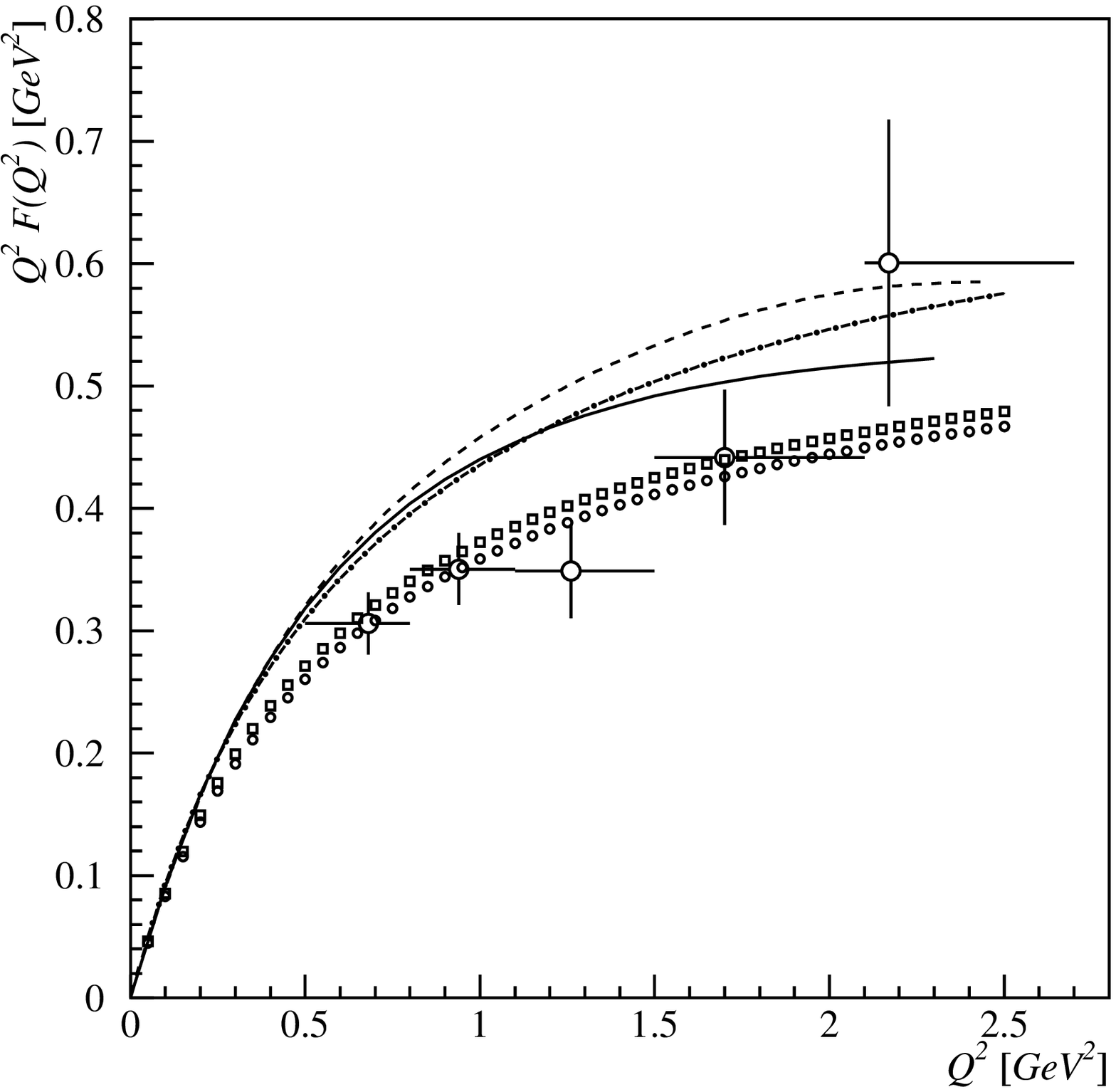}


\begin{thebibliography}{10}

\bibitem{bando94}
M. Bando, M. Harada, and T. Kugo, Prog. Theor. Phys {91} (1994) 927.

\bibitem{Roberts}
C.~D. Roberts, ``Electromagnetic Pion Form Factor and Neutral Pion Decay
  Width'', ANL preprint No. ANL--PHY--7842--94 (1994),
  to appear in Nucl. Phys. A; \\
  C. D. Roberts, in: Chiral Dynamics: Theory and Experiment,
  eds. A. M. Bernstein and B. R. Holstein, 
  Lecture Notes in Physics, Vol. {452} (Springer, Berlin, 1995) p. 68.

\bibitem{Frank+al}
M. R. Frank, K. L. Mitchell, C. D. Roberts and P. C. Tandy, Phys. Lett. B
  {359} (1995) 17.

\bibitem{Burden+al}
C. J. Burden, C. D. Roberts and M. J. Thompson,
Phys. Lett. B {371} (1996) 163.

\bibitem{Miransky}
V.A. Miransky, ``Dynamical Symmetry Breaking
 in Quantum field Theories", World Scientific Publishing Co.,
 Singapore 1993, ISBN 981--02--1558--4.

\bibitem{JJ}
R. Jackiw and K. Johnson, Phys. Rev. D {8} (1973) 2386.

\bibitem{CLEO1995}
V. Savinov and R. Fulton, ``Measurements of the two photon widths of the
  charmonium states $\eta_{c}$, $\chi_{c0}$ and $\chi_{c2}$'',
contribution to
  the PHOTON95 conference, Sheffield (1995), hep-ex/9507006.


\bibitem{jain91}
P. Jain and H.~J. Munczek, Phys. Rev. D {44} (1991) 1873.

\bibitem{munczek92}
H.~J. Munczek and P. Jain, Phys. Rev. D {46} (1992)  438.

\bibitem{jain93b}
P. Jain and H.~J. Munczek, Phys. Rev. D {48} (1993)  5403.

\bibitem{RW}
C.D. Roberts and A.G. Williams, 
Prog. Part. Nucl. Phys. 33 (1994) 477.

\bibitem{itzykson80}
C. Itzykson and J.-B. Zuber, Quantum field theory (McGraw--Hill, Inc.,
  Singapore, 1980).

\bibitem{F}
M. R. Frank, Phys. Rev. C {51}, 987 (1995).

\bibitem{BC}
J. S. Ball and T.-W. Chiu, Phys. Rev. D {22} (1980) 2542.

\bibitem{horvat91}
R. Horvat, D. Kekez, D. Klabu\v{c}ar and D. Palle, Phys. Rev. D {44}
  (1991) 1585.

\bibitem{horvat95}
R. Horvat, D. Kekez, D. Palle and D. Klabu\v{c}ar, Z. Phys. C {68} (1995)
   303.

\bibitem{behrend91}
CELLO Collaboration, H.-J. Behrend {\it et al.}, Z. Phys. C
	{49} (1991) 401.

\bibitem{Jaus}
W. Jaus, Phys. Rev. D {44} (1991) 2851.

\bibitem{IBG}
H. Ito, W. W. Buck and F. Gross, Phys. Lett. B {287} (1992) 23.

\bibitem{BBC}
J. Bijnens, A. Bramon and F. Cornet, Phys. Rev. Lett. 61 (1988) 1453.

\bibitem{ABBC}
Ll. Ametller, J. Bijnens, A. Bramon and F. Cornet, Phys. Rev.D {45} (1992)
986.

\bibitem{BH} 
M. Bando and M. Harada, Phys. Rev. D {49} (1994) 6096.

\bibitem{BrodskyLepage}
G. P. Lepage and S. J. Brodsky, Phys. Rev. D {22} (1980) 2157;
S. J. Brodsky and G. P. Lepage, {\it ibid.} {24} (1981) 1808.

\bibitem{PDG95}
L. Montanet {\it et al.}, Phys. Rev. D {50} (1994) 1173 and 1995 off--year
  partial update for the 1996 edition available on the PDG WWW pages (URL:
  http://pdg.lbl.gov/).

\bibitem{AhmadyMendel}
M. R. Ahmady and R. R. Mendel, Phys. Rev. D {51} (1995) 141.

\bibitem{Muenz}
C. R. M\UMLAUT{u}nz, ``Two--photon decays of mesons in a relativistic quark
  model'', University of Bonn preprint No. TK--96--01 (1996),
  hep-ph/9601206, submitted to Phys. Rev. D.

\end{thebibliography}
\end{document}